%% file: main.tex
\documentclass[sigconf]{acmart}

\acmConference[ICSE 2022]{The 44th International Conference on Software Engineering}{May 21-29, 2022}{Pittsburgh, PA, USA}

\AtBeginDocument{%
  \providecommand\BibTeX{{%
    \normalfont B\kern-0.5em{\scshape i\kern-0.25em b}\kern-0.8em\TeX}}}

\copyrightyear{2022} 
\acmYear{2022} 
\setcopyright{rightsretained} 
\acmConference[ICSE '22]{44th International Conference on Software Engineering}{May 21--29, 2022}{Pittsburgh, PA, USA}
\acmBooktitle{44th International Conference on Software Engineering (ICSE '22), May 21--29, 2022, Pittsburgh, PA, USA}
\acmDOI{10.1145/3510003.3510049}
\acmISBN{978-1-4503-9221-1/22/05}

\usepackage[section]{easy-todo}

\usepackage{amssymb}
\usepackage{amsmath, amsfonts}
\usepackage{graphicx}
\usepackage{textcomp}
\usepackage{xcolor}
\usepackage{hyperref}
\usepackage{cleveref}
\usepackage{xspace}
\usepackage{paralist}
\usepackage{multirow}
\usepackage{subcaption}
\usepackage{tikz}
\usepackage{courier}
\usepackage{listings} 
\usepackage{balance}

\newcommand{\cbert}{{\sc Code\-BERT}\xspace}
\newcommand{\cxglue}{{\sc Code\-XGLUE}\xspace}
\newcommand{\gcbert}{{\sc Graph\-Code\-BERT}\xspace}
\newcommand{\etal}{\emph{et al.}\xspace}
\newcommand{\ie}{\emph{i.e.},\xspace}
\newcommand{\eg}{\emph{e.g.},\xspace}
\newcommand{\mlcbert}{${\mathcal Polyglot}${\sc Code\-BERT}\xspace}
\newcommand{\mlgcbert}{${\mathcal Polyglot}${\sc Graph\-Code\-BERT}\xspace}

\definecolor{Gray}{gray}{0.3}
\tikzstyle{mybox} = [draw=black, very thick, rectangle, rounded corners, inner ysep=5pt, inner xsep=5pt, fill=gray!20]

\newcommand{\takeaway}[2]{
    \smallskip
    \noindent
    \begin{tikzpicture}
        \node [mybox] (box){%
        \centering
        \begin{minipage}{.465\textwidth}
        \fontsize{8.8}{10}\selectfont
        \textbf{Finding #1}. #2
        \end{minipage}
        };
    \end{tikzpicture}%
}

\newcommand{\link}[1]{{\color{blue} #1}}

\begin{document}

\title{Multilingual training for Software Engineering}


\author{Toufique Ahmed}
\affiliation{%
  \institution{University of California, Davis}
  \city{Davis}
  \state{California}
  \country{USA}
  \postcode{95616}}
\email{tfahmed@ucdavis.edu}

\author{Premkumar Devanbu}
\affiliation{%
  \institution{University of California, Davis}
  \city{Davis}
  \state{California}
  \country{USA}
  \postcode{95616}}
\email{ptdevanbu@ucdavis.edu}

\renewcommand{\shortauthors}{Toufique Ahmed and Premkumar Devanbu}

\begin{abstract}



Well-trained machine-learning models, which leverage large amounts 
of open-source software data, 
have now become an interesting approach to automating many 
software engineering tasks. Several SE tasks  have
all been subject to this approach, with performance gradually improving over the
past several years with better
models and training methods. More, and more diverse,  \emph{clean, labeled} data is better for training; but constructing
good-quality datasets is time-consuming and challenging. 
Ways of augmenting
the volume and diversity of clean, labeled data generally have wide applicability. For some languages (\emph{e.g.,} Ruby) 
labeled data is less
abundant; in others (\emph{e.g.,} JavaScript) the available data maybe more focused on some application
domains, and thus less diverse. 
As a way around such data bottlenecks,
we present evidence suggesting that human-written code in different languages (which performs the same function), 
is rather similar, and particularly preserving  of identifier naming patterns; we further present
evidence suggesting that identifiers are a very important element of training data for software engineering tasks. 
We leverage this rather fortuitous phenomenon to 
find evidence that available \emph{multilingual} training data (across
different languages)  can be used to amplify performance. 
We study this for 3 different tasks: code summarization, code retrieval, and function naming.  
We note that this data-augmenting approach is broadly compatible
with different tasks, languages, and machine-learning models. 
\end{abstract}

%
%

\ccsdesc[500]{Software and its engineering~Software notations and tools}
\ccsdesc[500]{Computing methodologies~Machine learning}
\keywords{code summarization, code search, method name prediction, deep learning}
\maketitle

\input{introduction.tex}
\input{background.tex}

\input{task_dataset.tex}

\input{result.tex}
\input{interpret.tex}

\input{related_work.tex}
\input{conclusion.tex}



\balance
\bibliographystyle{ACM-Reference-Format}
\bibliography{acm}

\end{document}

%% file: introduction.tex
\section{Introduction}

Researchers  in the NLP  area have reported that 
multilingual training is  beneficial for low-resource language~\cite{dabre2020survey,ha2016toward,ranathunga2021neural,tang2020multilingual}. Several papers show that multilingual-trained models show better performance~\cite{tan2019multilingual,lakew2018comparison} and are more practical to deploy~\cite{arivazhagan2019massively}. However, this is observed in two situations: 1) for low-resource languages and 2) when the languages
are related. We find that programs in different languages solving the same problem use more similar identifiers; 
furthermore different languages sometimes have similar keywords and operators. High capacity deep learning models are capable of learning \textit{interlingua}: shared semantic representation between languages~\cite{johnson2017google}. Moreover, with tasks like summarization, or method naming,  
we are dealing with a simplified, many-to-one  setting: translating multiple source languages to a single target language), which is believed to be easier than multi-way task~\cite{garmash2016ensemble,zoph2016multi}. We begin by introducing the code summarization task, which we use to motivate multilingual training.  

Developers often rely heavily on comments, to gain a quick
(even if approximate) understanding of the specification and design of code
they are working on. An actual example of a comment
is shown in Figure~\ref{fig-ex}.  
Such comments help a developer gain a quick mental preview of \emph{what} the proximate code  does, and \emph{how}
it might go about it; this helps the developer know what to look for in the code. 
Knowing that such comments are useful to others (or even later to oneself) incentivizes developers to  create
comments that explain the code; however the resulting  redundancy (\emph{viz.,} code that does something,
and some nearby English text that describes just what the code does),  with the same concept expressed in two  languages
results in a bit of extra work for the original coder. 


\noindent 
This extra work, of creating aligned comments explaining the code, can be fruitfully viewed~\cite{gros2020code} as a task related to \emph{natural language
translation} (NLT) (\emph{e.g.,} translating English to German). The mature \& powerful technology of NLT
becomes applicable for comment synthesis; ML approaches developed for the former
can be used for the latter. An effective comment synthesizer could help developers: by saving them 
the trouble of writing comments;  and perhaps even be used on-demand in the IDE to create descriptions of selected 
bits of code. 

Comment synthesis is now an active research area, including many projects such as CodeNN~\cite{iyer2016summarizing}, DeepCom~\cite{hu2018deep}, Astattgru~\cite{leclair2019neural}, \cbert~\cite{feng2020codebert}, Rencos~\cite{zhang2020retrieval}, SecNN~\cite{li2021secnn}, PLBART~\cite{ahmad-etal-2021-unified}, CoTexT~\cite{phan2021cotext}, ProphetNet-X~\cite{qi2021prophetnet}, NCS~\cite{ahmad2020summarization}, Code2seq~\cite{alon2018codeseq}, Re$^2$Com~\cite{wei2020retrieve}, and many more~\cite{leclair2021ensemble,haque2020improved,gao2021code,leclair2020improved,wei2019code,
wan2018improving,hu2018summarizing,hu2020deep,li2020deepcommenter,wang2020reinforcement,yuchao2021yet,
yang2021comformer,wang2021cocosum,mastropaolo2021studying,hasan2021codesc,mahmud2021code}.   
All these approaches rely on datasets of aligned code-comment pairs. 
Typically, these datasets are then used 
to train complex deep learning models to model a probabilistic distribution of the form {\small\em p(comments | code) }; one can 
sample from these (usually generative) models to create candidate comments for a given a piece of code. Given a dataset of code-comment pairs
in  a  specific language, \emph{e.g.,} Java, or Python, or PHP, or Ruby, one can train models to translate code in \emph{that} language
to comments. The quality of the translation will depend largely upon the inductive power of the model, and quality and diversity
of the code-comment dataset.

\begin{figure}[htb]
\captionsetup{aboveskip=-0pt,belowskip=0pt}
\vspace{-0.05in}
\centering
\begin{lstlisting}[language=Java,basicstyle=\scriptsize]
//Returns the text content of 
//this node and its descendants.
public String getTextContent() {
    StringBuilder sb=new StringBuilder(getChildNodesCount()+1);
    appendTextContent(sb);
    return sb.toString();
}
\end{lstlisting}
\caption{{\em {\small Example for code comment generation task}}}
\label{fig-ex}
\end{figure}

Of late, given the power of GPUs, and the capacity of the models, the limitations largely arise from 
dataset quality and diversity, especially in languages for which limited, or rather specialized data is available. For instance, \cxglue~\cite{DBLP:journals/corr/abs-2102-04664} dataset consists of six languages (\ie Ruby, Java, JavaScript, Go, Php, Python). Most languages have well over 100,000 training examples, 
covering a wide set of application domains. Some languages, particularly
 Ruby and Javascript, have far fewer examples, and cover a narrower range of
 application domains.  
As a result,  state-of-the-art models perform less well for these two languages.  This is a well-known
problem for natural language translation:  while training data for language pairs like $English \leftrightarrow French$
is abundant, resources may be lacking for less-used languages
like $Quechua$ or $Badaga$. In such cases, a common technique is adapt ML models to learn useful
statistics from abundant data in  other, perhaps related languages~\cite{nakov2009improved}. This works
well when  languages often have similar grammars, and  share common word etymologies. 


We propose an analogous approach to improve the diversity and quality of
training data for software-engineering tasks,  
exploiting an interesting property of source
code that human beings write. 
It's generally agreed that variable names help code comprehension~\cite{lawrie2007effective}. Developers
know this, and typically choose  descriptive variable names (reflective  of code logic
and purpose) regardless
of the language they are coding in. Thus, one could expect that developers coding \emph{the same functionality,
using similar algorithms, even in different languages, will use similar variable names}. This suggests that machine-learning 
approaches could sometimes leverage corpora in different programming languages. This paper a) shows that  this  expectation actually 
has a sound empirical basis, and then b) demonstrates that this approach in fact works not
just for code summarization, but also for several other tasks. 
We make the following contributions. 

\begin{enumerate}
\item Using the {\small\sc RosettaCode} dataset, we provide evidence that programs solving the same problem 
in different languages are more likely to use the same or similar identifier names. 
\item We show evidence suggesting that cross-language training (\emph{e.g.,} train on Python, test on Ruby)
can sometimes lead to better performance than same-language training. 
\item We study the relative value of identifiers and syntax, using ablation, and find that identifier names may matter more. 
\item We show that pooled  multilingual training data improves performance
on several tasks, but especially for languages
lacking in diverse and abundant data. We top a leaderboard for code-comment synthesis\footnote{This claim is based
on publicly available evidence. Please check \textit{https://microsoft.github.io/CodeXGLUE/}}.
\item We show that multilingual training helps for two other tasks: \emph{code retrieval}, and \emph{method name prediction}. 
\item Finally, we evaluate a few different design choices for multilingual training, and discuss threats to our findings.  
\end{enumerate}

Overall, this paper a) shows that multilingual training is yet another useful technique in the general arsenal of ML approaches
to exploit the naturalness of code, b) shows why it is useful, and c) shows how to take good advantage of it. 

\smallskip

\noindent{\bf Note:} Technical details follow, but precisely: what  we study here is  multilingual training in 
the \emph{fine-tuning stage of ``foundation models"~\cite{bommasani2021opportunities}}. 
 Foundation models for code,  like \cbert, \gcbert~\cite{feng2020codebert, guo2020graphcodebert} already use multilingual
data for \emph{pre-training}. While pre-training is  self-supervised and is
done with unlabeled corpora, task-specific fine-tuning is usually supervised, using clean, hard-won labeled data; multilingual pooling
can be  useful  here. 


%% file: background.tex
\section{Background \& Motivation}
We now present some motivating evidence suggesting the value of  multilingual training data for
deep-learning applications to software tasks. We begin the argument focused on code summarization. 

Deep learning models have been widely applied to code summarization, 
with papers reporting substantial gains in  performance over recent years~\cite{iyer2016summarizing,hu2018deep,leclair2019neural,feng2020codebert,zhang2020retrieval,
li2021secnn,ahmad-etal-2021-unified,phan2021cotext,qi2021prophetnet,ahmad2020summarization,alon2018codeseq,wei2020retrieve,leclair2021ensemble,haque2020improved,gao2021code,leclair2020improved,wei2019code,
wan2018improving,hu2018summarizing,hu2020deep,li2020deepcommenter,wang2020reinforcement,yuchao2021yet,
yang2021comformer,wang2021cocosum,mastropaolo2021studying,hasan2021codesc}.  
We focus here on \emph{what information in the code} ML models leverage
 for summarization (while we use summarization to motivate the approach, we evaluate later on 3 different tasks). 
Does every token in the program under consideration matter, for the code summarization task? Or, are the function and variable names used in the programs 
most important? Since identifiers carry much information about the program, this
may be a reasonable assumption. 

Considering the content words\footnote{``Content'' words  in linguistics, are words that carry meaning, as contrasted with
\emph{function} words, such as prepositions, pronouns, and conjunctions, which denote grammatical relationships. See \url{https://en.wikipedia.org/wiki/Content_word}. In code, we consider function words to be keywords, operators and punctuations, and content words
to be identifiers (functions, variables, types, \emph{etc})}
 in the example in Figure~\ref{fig-ex}  there are four major terms (\ie Returns, text content, node, and descendants) used in the summary. The first 3 directly occur as tokens or subtokens in the code. Though the word ``descendants" is missing in the program, high capacity neural models like BERT~\cite{devlin2018bert} can learn to statistically connect, \emph{e.g.,} "descendant" with the identifier subtoken ``child''. 
 This suggests that, perhaps, comments are recoverable primarily from identifiers.  If this is so, and 
 identifiers matter more for comments than the exact syntax of the
 programming language, that may actually be very good news indeed. If developers choose identifiers in the same way across different languages
 (\emph{viz.,} problem-dependent, rather than language dependent)
perhaps we can  improve the diversity and quality of dataset by pooling training set across may languages. 
Pooled data sets may allow us to fine-tune using multilingual data, and improve performance, especially for low-resource languages (\eg Ruby and JavaScript from \cxglue~\cite{DBLP:journals/corr/abs-2102-04664}).  
Since this is a core theoretical background for our work, 
we start off with two basic research questions to empirically gauge the possibility and promise of multilingual fine-tuning.

\begin{enumerate}
\item[{\bf RQ1}] What role do identifiers play in for code summarization? 
\item[{\bf RQ2}] Do programs that solve the same problem 
in different languages tend to use similar identifier names?    

\end{enumerate}
\subsection{RQ1: Role played by identifiers}

We first examine the importance of \emph{identifiers} for  code summarization; specifically, 
we compare the relative value of identifier tokens and other tokens. 
 We use the  \cxglue dataset and pre-trained \cbert  embeddings for the task~\cite{feng2020codebert}.  We begin with a brief backgrounder on \cbert~\cite{feng2020codebert} \& BERT~\cite{devlin2018bert}. 

\cbert uses the pre-training + fine-tuning strategy of BERT, RoBERTa \emph{etc}~\cite{devlin2018bert,liu2019roberta}.  This approach begins with a self-supervised 
 ``pre-training''  step, to learn textual patterns from a large, unlabeled, corpus using just the content;
 in the next step, ``fine-tuning'',   task-specific \emph{labeled} data is used to provide task-related
 supervised training. This approach is known 
 to achieve state-of-the-art performance
 in both natural language processing, and software-related tasks~\cite{feng2020codebert,jesse2021learning,guo2020graphcodebert,zhang2020sentiment,biswas2020achieving,
 9520296,kanade2019pre,jiang2021cure,ahmed2021synfix,mastropaolo2021studying}. 

We study the effect of identifiers in several steps. For the pre-training step, we start with  the available \cbert model, which
is pre-trained on a large, multilingual corpus of code.  
For the fine-tuning step, for this task, we use the \cxglue benchmark dataset (see \cref{dataset} for languages and dataset sizes); 
we start with the original set of code-comment pairs, and apply two different treatments to create overall three different 
fine-tuning training datasets--1) base case leaving code as is, 2) a treatment to emphasize identifiers, and 3) a treatment to de-emphasize them. 
First, to emphasize identifiers  
we abstract out the program's keywords, separators, and operators by replacing those with three generic tokens (\ie ``key'', ``sep'', and ``opt''), thus
forcing the model (during fine-tuning) to rely more on the identifiers, for the task. 
Next, to assess the importance of keywords, separators, and operators,
we abstract out the identifiers with a generic token ``id''. We fine-tune the model separately after each of these abstraction steps, thus yielding 3 fine-tuned
models: the baseline, keyword-abstracted, and identifier-abstracted. We compare the results (smoothed BLEU-4)  across all three. 

If a fine-tuned model's performance is relatively unaffected by an abstraction, one may infer that the model relies less on the abstracted tokens. We perform these experiments with two languages with low-resource  (\ie Ruby and JavaScript, See \cref{dataset}) and two languages with high-resource (\ie Java and Python ). We train, validate, and test with the same dataset in each case. For each test instance, we have one value from the complete program and another one from each of the two abstracted versions. 
We compared these values, using two distinct pair-wise Wilcoxon tests:
1) Alternative Hypothesis (AH): complete program > identifier de-emphasis \& 2) AH: complete program > identifier emphasis. 
We also perform the same test with the keyword-abstracted and identifier-abstracted versions (AH: identifier emphasis > identifier de-emphasis). 


\begin{table}[h]

\centering
\resizebox{\columnwidth}{!}{%
\renewcommand{\arraystretch}{1.2}
\begin{tabular}{llllllll}
\hline
\multicolumn{1}{c}{\multirow{2}{*}{Dataset}} & \multicolumn{1}{c}{\begin{tabular}[c]{@{}c@{}}Complete\\ Program\end{tabular}} & \multicolumn{3}{c}{\begin{tabular}[c]{@{}c@{}}Abstracting keyword, \\ operator, separator\end{tabular}}                                                                                     & \multicolumn{3}{c}{\begin{tabular}[c]{@{}c@{}}Abstracting\\ identifiers\end{tabular}}                                                                                                       \\ 
\multicolumn{1}{c}{}                         & \multicolumn{1}{c}{BLEU-4}                                                     & \multicolumn{1}{c}{BLEU-4} & \multicolumn{1}{c}{\begin{tabular}[c]{@{}c@{}}Effect\\ Size\end{tabular}} & \multicolumn{1}{c}{\begin{tabular}[c]{@{}c@{}}p-value\\ (adjusted)\end{tabular}} & \multicolumn{1}{c}{BLEU-4} & \multicolumn{1}{c}{\begin{tabular}[c]{@{}c@{}}Effect\\ Size\end{tabular}} & \multicolumn{1}{c}{\begin{tabular}[c]{@{}c@{}}p-value\\ (adjusted)\end{tabular}} \\ \hline
Ruby                                           & \textbf{12.53                                                                          } & 11.57                       &  -0.028      & 0.008 & 7.94                        & -0.238                                                                      & \textless{}0.001                                                                  \\
JavaScript                                     & \textbf{13.86}                                                                           & 13.06                       & -0.033       & \textless{}0.001                                                                  & 9.06                        & -0.175 & \textless{}0.001                                                                  \\

Java                                           & \textbf{18.72} & 18.72                       & -0.002                                                                      & 0.344                                                                            & 11.41                       & -0.254                                                                      & 0                                                                                 \\ 

\multicolumn{1}{l}{Python}                   & \multicolumn{1}{l}{\textbf{18.25}}                                                      & \multicolumn{1}{l}{18.10}  & \multicolumn{1}{l}{-0.010}                                                 & \multicolumn{1}{l}{\textless{}0.001}                                             & \multicolumn{1}{l}{11.68}  & \multicolumn{1}{l}{-0.288}                                                 & \multicolumn{1}{l}{0}                                                            \\ \hline
\end{tabular}
}
\vspace{0.05in}
\caption{{\em {\small Role played by identifiers}}}
\vspace{-0.2in}
\label{role}
\end{table}

The data (\cref{role}) suggests that abstracting the keyword, separator, and operator has a smaller  impact on the performance: the BLEU-4 scores 
are rather similar (with \emph{effect size} ranging from 0.002 to 0.033)  to those from the unabstracted code. On the other hand, when de-emphasizing identifiers, the performance drops more, with effect sizes 5x-100x  larger. We find similar results while comparing the emphasizing and de-emphasizing identifiers versions (omitted for brevity).

\begin{table}[h]

\centering
\resizebox{\columnwidth}{!}{%
\renewcommand{\arraystretch}{1.2}
\begin{tabular}{llllllll}
\hline
\multicolumn{2}{c}{\multirow{2}{*}{Language}} & \multicolumn{6}{c}{Training}                                                                                                                                      \\ 
\multicolumn{2}{c}{}                          & \multicolumn{1}{c}{Ruby} & \multicolumn{1}{c}{JavaScript} & \multicolumn{1}{c}{Java} & \multicolumn{1}{c}{Go} & \multicolumn{1}{c}{PHP} & \multicolumn{1}{c}{Python} \\ \hline
\multirow{6}{*}{Testing}        & Ruby          & \textit{12.53}                     & 11.84                   & \textbf{13.42}                     & 12.32                   & \textbf{13.84}                    & \textbf{14.09}                       \\
                                & JavaScript            & 11.98                     & \textit{13.86}                   & \textbf{14.16}                     & 12.55                   & \textbf{13.90}                    & \textbf{14.09}                       \\
                                & Java          & 13.38                     & 14.57                   & \textit{\textbf{18.72}}                     & 14.20                    & 16.27                    & 16.20                       \\
                                & Go            & 11.68                     & 11.24                   & 13.61                     & \textit{\textbf{18.15}}                   & 12.70                    & 13.53                       \\
                                & PHP           & 17.52                     & 19.95                   & 22.11                     & 18.67                   & \textit{\textbf{25.48}}                    & 21.65                       \\
                                & Python        & 14.10                     & 14.44                   & 16.77                     & 14.92                   & 16.41                    & \textit{\textbf{18.25                    } } \\ \hline
\end{tabular}
}
\vspace{0.05in}
\caption{{\em {\small Intra and inter language training and testing}}}
\label{tab_intra}
\vspace{-0.2in}
\end{table}

The results in \cref{role}
suggests that syntax is less relevant that identifier names. 
 In all the prior works, the training and testing were done in the same language. Since syntax is less important,  
could we train and test with different languages? The \cxglue dataset enables just such an experiment. Using six different languages, 
we apply a \cbert model  fine-tuned in each language, 
to a test set in another language. 
Table~\ref{tab_intra} shows that for high-resource languages (\ie Java, go, PHP, and Python), we achieve the best result  (diagonal)
when training and test data are from the same language.  However, the performance does not degrade 
to a very large extent when trained with one language and tested on a different one. Surprisingly we observe that for Ruby and JavaScript, 
we actually \emph{achieve higher performance while trained with Java, PHP, and Python than the language itself}. That indicates that code summarization
is not completely dependent on syntax (perhaps it relies more on identifier similarity, which we shall explore next)

\takeaway{1}{Code summarization sometimes appears to train quite well with data sets from other languages, even if the syntax is different.}

\subsection{RQ2: Identifier similarity across Languages}

Here, we evaluate RQ2: given a problem, do developers choose similar, descriptive identifiers, regardless of the programming language? Based
on the findings in the previous section: if identifiers were indeed used in similar ways, 
perhaps code-comment pairs \emph{from any programming language} could help train a code summarization model,  
\emph{for any other language}. As an example, 
Figure~\ref{figindex} presents that all the ``indexOf'' functions implemented in Java, PHP and JavaScript  use very similar identifiers ``needle'' and ``haystack''. 

Quantitatively evaluating this hypothesis requires multiple implementations of the same problem in different programming languages, where we could compare
identifier names.  
Luckily, {\small\sc RosettaCode} provides just such a dataset. {\small\sc RosettaCode} currently consists of 1,110 tasks, 305 draft tasks and includes 838 languages\footnote{Last Accessed August, 2021}. We collect the mined data\footnote{https://github.com/acmeism/RosettaCodeData} and study the same six languages (\ie Ruby, JavaScript, Java, Go, PHP, and Python) in the \cxglue dataset. We get 15 cross-language pairs from six languages and measure identifier similarity between pairs of programs
which solve the same problem in each language (\emph{e.g.,} programs for graph diameter problem in Java and Ruby). 
For baselining, we also compare with  a random pair (solving different problems) for the same two languages (\emph{e.g.} graph diameter in Java, and
SHA-hashing in Ruby). Fortunately, we found sufficient sample sizes for all our language pairs in {\small\sc RosettaCode}. 
For example, for Java \& Python we find 544 matched program pairs solving the same problem in both languages. We then take the  544 Java programs and randomly pair them with 544 other Python programs. Therefore, we have two groups of programs (\ie same program implemented in different languages and different programs implemented in different languages), and we check the similarity level between the two groups.  
Note that \emph{size-unrestricted} random pairing may yield misleading results. Suppose we have a Java \& Python program \emph{matched} pair with 100 Java subtokens and 40 Python subtokens. Now, if we replace the matched python program with a random, bigger program (\eg 500 subtokens), we may have more chance of finding matched identifiers. 
Therefore, while choosing the random program, we try to ensure it has a similar length to the program it is replacing in the pair. We randomly select a program having the subtoken counts within a 5\% length range (\eg 38-42 subtokens for a 40 subtoken program) of the removed one. Fortunately, in \emph{99.25\%} cases, we get at least one example within the 5\% range. On the remaining instances, we select the program with the nearest subtoken count.

We measure identifier similarity thus: 
\begin{enumerate}
\item Remove all keywords, operators, and separators from the programs.
\item Break all CamelCase and snake\_case identifiers and keep only one copy of each sub token.
\item Discard too-small programs with less than 5 sub-tokens. 
\item Calculate the mean Szymkiewicz-Simpson coefficient (overlap coefficient)~\cite{vijaymeena2016survey} for both groups (\ie same program pair and random pair) of programs.
\item Repeat this process across all 15 language pairs, for all program pairs. 
\end{enumerate}

Table~\ref{tab_sim} shows the common program pairs have 89\%-235\% \emph{additional {\bf identifier} overlap }
compared to random program pairs. We compare
the matched and random pair overlaps using the non-parametric  Wilcoxon signed-rank test (AH: random has less overlap than matched). 
We observe that the null hypothesis is rejected, and Szymkiewicz-Simpson Overlap coefficient\footnote{This is a measure of similarity like the Jaccard index; we use it here since sometimes the sizes of the programs are quite different. It's calculated as  $\frac{\mid X \cap Y \mid}{min ( \mid X \mid, \mid Y \mid)}$. }
is significantly higher for the common program pairs in all the cases. That indicates programs solving  the same problem (even in different languages) are much more likely to use the same or similar identifier names.

\begin{table}[h]

\centering
\resizebox{\columnwidth}{!}{%
\renewcommand{\arraystretch}{1.2}
\begin{tabular}{lllllll}
\hline
\multicolumn{1}{c}{\multirow{2}{*}{\begin{tabular}[c]{@{}c@{}}Language \\                 pair\end{tabular}}} & \multicolumn{1}{c}{\multirow{2}{*}{\begin{tabular}[c]{@{}c@{}}\#of common \\ programs\end{tabular}}} & \multicolumn{3}{c}{Overlap coefficient}                                                                                                                                                             & \multicolumn{1}{c}{\multirow{2}{*}{\begin{tabular}[c]{@{}c@{}}Effect \\ Size\end{tabular}}} & \multicolumn{1}{c}{\multirow{2}{*}{\begin{tabular}[c]{@{}c@{}}p-value\\  (adjusted)\end{tabular}}} \\ 
\multicolumn{1}{c}{}                                                                                          & \multicolumn{1}{c}{}                                                                                 & \multicolumn{1}{c}{\begin{tabular}[c]{@{}c@{}}for random\\  programs\end{tabular}} & \multicolumn{1}{c}{\begin{tabular}[c]{@{}c@{}}for common \\ programs\end{tabular}} & \multicolumn{1}{c}{increased in \%} & \multicolumn{1}{c}{}                   & \multicolumn{1}{c}{}                                                                               \\ \hline

Java \& Python                                                                                                  & 544                                                                                                   & 0.10                                                                                & \textbf{0.32                                                                               } & +210.67\%                            & 0.747                                 & \textless{}0.001                                                                   \\

Java \& Ruby                                                                                                    & 532                                                                                                   & 0.11                                                                               & \textbf{0.31                                                                               } & +174.97\%                            & 0.751                                 & \textless{}0.001                                                                   \\

Java \& Javascript                                                                                              & 411                                                                                                   & 0.13                                                                               & \textbf{0.36                                                                               } & +188.17\%                            &  0.774                                   & \textless{}0.001                                                                   \\

Java \& Go                                                                                                      & 602                                                                                                   & 0.19                                                                                & \textbf{0.36                                                                               } & +89.24\%                             & 0.641                                 & \textless{}0.001                                                                   \\

Java \& PHP                                                                                                     & 282                                                                                                   & 0.08                                                                                & \textbf{0.28                                                                               } & +235.01\%                            &  0.740                                  & \textless{}0.001                                                                                                                                                          \\

Python \& Ruby                                                                                                  & 538                                                                                                   &  0.11                                                                              & \textbf{0.35                                                                               } & +228.89\%                            & 0.780                                 & \textless{}0.001                                                                                                                                                          \\

Python \& Javascript                                                                                            & 377                                                                                                   & 0.12                                                                              & \textbf{0.34                                                                               } & +190.09\%                            & 0.728                                   & \textless{}0.001                                                                   \\

Python \& Go                                                                                                    & 601                                                                                                   & 0.13                                                                                & \textbf{0.31                                                                               } & +133.06\%                            & 0.664                                  & \textless{}0.001                                                                   \\

Python \& PHP                                                                                                   & 267                                                                                                   & 0.09                                                                               & \textbf{0.29                                                                               } & +214.32\%                            & 0.679                                   & \textless{}0.001                                                                                                                                                          \\

Ruby \& Javascript                                                                                              & 370                                                                                                   & 0.13                                                                               & \textbf{0.35                                                                               } & +167.02\%                            &  0.751                                  & \textless{}0.001                                                                   \\

Ruby \& Go                                                                                                      & 571                                                                                                   & 0.12                                                                              & \textbf{0.28                                                                               } & +133.47\%                            & 0.724                                 & \textless{}0.001                                                                   \\

Ruby \& PHP                                                                                                     & 262                                                                                                   & 0.09                                                                               & \textbf{0.28                                                                               } & +205.32\%                            & 0.716                                 & \textless{}0.001                                                                   \\

Javascript \& Go                                                                                                & 418                                                                                                   & 0.14                                                                               & \textbf{0.29                                                                               } & +110.96\%                             &  0.635                                  & \textless{}0.001                                                                                                                                                          \\

Javascript \& PHP                                                                                               & 236                                                                                                   & 0.11                                                                                & \textbf{0.29                                                                               } & +175.03\%                            & 0.678                                 & \textless{}0.001                                                                                                                                                          \\

Go \& PHP                                                                                                       & 293                                                                                                   & 0.10                                                                                & \textbf{0.23                                                                               } & +121.25\%                            & 0.562                                 & \textless{}0.001                                                                   \\ \hline
\multicolumn{1}{l}{Overall}                                                                                   & \multicolumn{1}{l}{6304}                                                                             & \multicolumn{1}{l}{0.12}                                                           & \multicolumn{1}{l}{\textbf{0.31}}                                                           & \multicolumn{1}{l}{+158.94\%}       & \multicolumn{1}{l}{0.697}           & \multicolumn{1}{l}{0}                                                                              \\ \hline
\end{tabular}
}
\vspace{0.05in}
\caption{{\em {\small Cross-language identifier similarity, when functionality is preserved}}}
\label{tab_sim}
\vspace{-0.2in}
\end{table}

We also calculate each pair's Jaccard index~\cite{jaccard1901etude} (similarity coefficient) and find 112\%-309\% more similarity between common pairs than random ones,
thus, giving essentially the same result.  However, we prefer to report the detailed result using the overlap coefficient because Jaccard index can be affected by the differing verbosity of languages. For example, on average, Java, Python, and Ruby programs in {\small\sc RosettaCode} have 29.45, 17.93, and 17.63 identifier subtokens. Java has higher subtokens compared to Python and Ruby because of the import statements, package naming etc. Therefore, Jaccard index between  Java and Python will be lower than that of Python and Ruby even if the programs use very similar identifiers. 

\begin{figure}[htb]
\captionsetup{aboveskip=-0pt,belowskip=0pt}
\centering
\begin{subfigure}[t]{0.53\textwidth}

\begin{lstlisting}[language=Java,basicstyle=\scriptsize]

public static int indexOf(ByteBuf needle, ByteBuf haystack) {
  // TODO: maybe use Boyer Moore for efficiency.
  int attempts = haystack.readableBytes() - needle.readableBytes() + 1;
  for (int i = 0; i < attempts; i++) {
    if (equals(needle, needle.readerIndex(),
      haystack, haystack.readerIndex() + i,
      needle.readableBytes())) {
      return haystack.readerIndex() + i;
    }
  }
  return -1;
}
\end{lstlisting} 
\vspace{-0.1in}
\caption{{\em {\small Java}}}
\end{subfigure}

\begin{subfigure}[t]{0.53\textwidth}

\begin{lstlisting}[language=Java,basicstyle=\scriptsize]

public static function indexOf(string $haystack, string $needle,
	 int $offset=0):int
{
  $pos=self::strpos($haystack, $needle, $offset);
  return is_int($pos)?$pos:-1;
}
\end{lstlisting}
\vspace{-0.1in}
\caption{{\em {\small PHP}}}
\end{subfigure}

\begin{subfigure}[t]{0.53\textwidth}

\begin{lstlisting}[language=Java,basicstyle=\scriptsize]
function indexOf(haystack, needle) {
  if (typeof haystack==='string')
    return haystack.indexOf(needle);
  for (let i=0, j=0, l=haystack.length, n=needle.length; i<l; i++) {
    if (haystack[i]===needle[j]) {
      j++;
      if (j===n) return i-j+1;
    }
    else {
      j=0;
    }
  }
  return -1;
}
\end{lstlisting}
\vspace{-0.08in}
\caption{{\em {\small JavaScript}}} 
\vspace{0.02in} 
\end{subfigure}

\vspace{0.07in}
\caption{{\em {\small Usage of similar identifiers (\eg needle, haystack) in ``indexOf'' function in different programming languages}}}
\label{figindex}
\vspace{-0.15in}
\end{figure} 

\takeaway{2}{For a given problem, developers are likely to choose similar identifiers, even if coding in different languages.}

In this section, we have presented evidence suggesting that a) identifiers are important for code summarization, that
b) cross-language training is promising, and also that c) identifiers tend to be used in similar ways across languages. 
Taken together, these findings present a strong argument to try multilingual fine-tuning for SE tasks. 
Note that it is already well established that multi-lingual pre-training is helpful, and most BERT-style SE pre-trained models are multilingual~\cite{feng2020codebert, ahmad-etal-2021-unified,phan2021cotext,qi2021prophetnet}. However, pre-training data are unsupervised and easy to collect. Preparing clean data for the  supervised  fine-tuning  phase
requires more time and attention.  In this paper, our aim is to prove that multilingual training is not only effective in pre-training stage but also in fine-tuning stage for SE models, which is already found to be beneficial for \textit{natural language} models~\cite{tang2020multilingual}. 


%% file: task_dataset.tex
\section{Benchmark Datasets and Tasks}
\label{task}
We evaluate the benefits of multilingual training in the context of several tasks, and associated datasets. 
In this section, we discuss the models and tasks used for our experiments.

\subsection{The Models}
\label{cgcmodels}
For our study of multilingual training, we adopt the BERT, or ``foundation model'' paradigm.  Foundation models~\cite{devlin2018bert,liu2019roberta,conneau2019cross,raffel2019exploring,brown2020language} have two stages:  i) unsupervised pre-training with  corpora at vast scale and ii) fine-tuning with a smaller volume of supervised data for the actual task. Foundation models currently hold
state-of-the-art performance for a great many NLP tasks. 
BERT~\cite{devlin2018bert} style models have
also been adapted for code, pre-trained on a huge, multilingual,  corpora, and made  available: \cbert and \gcbert  
are both freely available: both source  code and pre-trained model parameters. 
 While these models  
for code have thus far generally been fine-tuned monolingually, they provide an excellent platform for
training experiments like ours, to measure the gains of multilingual fine-tuning. 
 \cbert \& \gcbert use a multi-layer bidirectional Transformer-based~\cite{vaswani2017attention} architecture, and it is exactly as same as the RoBERTa~\cite{liu2019roberta}, with 125M parameters; we explain them further below.

\noindent{\underline{\em Pre-training} }
The \emph{\cbert}~\cite{feng2020codebert} dataset, 
has two parts: a matched-pairs part with 2.1M pairs of function and associated comment (NL-PL pairs) 
and 6.4M samples with just code. The code includes several programming languages. It was created by Hussain \etal~\cite{husain2019codesearchnet}. 
\cbert model is pre-trained with two objectives (\ie Masked Language Modeling and Replaced Token Detection) on both parts. 
Mask language Modeling (MLM) is a widely applied and effective~\cite{devlin2018bert,liu2019roberta} training objective where a certain number of (15\%) tokens are masked out, and the model is asked to find those tokens. For \cbert training, Feng \etal apply this first objective only to bimodal data~\cite{feng2020codebert}.
The second objective, Replaced Token Detection (RTD)~\cite{clark2020electra}, is a binary classification problem that is applied to both unimodal and bimodal data. Two data generators (\ie NL and PL) generate plausible alternatives for a set of randomly masked positions, and a discriminator is trained to determine whether a word is the original one or not.
We note that \cbert pre-training is all about representation-learning: by learning to perform the task well, the model learns a good way to \emph{encode} the
text, which is helpful during the next, fine-tuning stage. The pre-training took about 12 hours on a machine with 16 NVIDIA V100 cards,
and would have taken us very much longer, so we were grateful to be able to just download the estimated parameters.

\noindent{\underline{\em Pre-training \gcbert}}
\gcbert augments source-code with  data flow, during pre-training.  It uses a simple
data flow graph (DFG)  encoding a \emph{where-the-value-comes-from} relation between variables~\cite{guo2020graphcodebert}.
The DFG nodes are variable occurrences, edges are value flow. 
\gcbert pretraining learns a joint representation  of 1) the DFG  structure, 2)  DFG alignment
with source code, and 3) the source code token sequences. 
 \gcbert is therefore pre-trained with three training objectives (\ie Edge Prediction, Node Alignment, and MLM) on 2.3M functions (PL-NL pairs) from CodeSearchNet~\cite{husain2019codesearchnet} dataset. For details see~\cite{guo2020graphcodebert}.

The pre-training+fine-tuning approach relies on VERY high capacity models, and are pre-trained over a large,  multilingual corpus. Thus, even before fine-tuning, the models already know a lot about each language. Thus, fine-tuning on many languages should not negatively impact what the model knows about any one language. Thus we find that multilingual fine-tuning  improves on monolingual fine-tuning in most cases. We believe our proposed approach would still consider the context surrounding the individual programming language even after multilingual training because these models have sufficient capacity to do so.

 We now describe our tasks: in each, we describe the task, the dataset, and the multilingual fine-tuning approach (if applicable). 

\subsection{Code Summarization}
\label{codesum}

\noindent{\em \underline{The Task: }} as described earlier, the goal is to generate a NL summary given code in some PL.

\noindent{\em \underline{\em The Dataset: }}
There are several different code summarization datasets; we chose \cxglue\footnote{CodeSearchNet~\cite{husain2019codesearchnet} dataset is a standard benchmark, which has
been incorporated into \cxglue}~\cite{DBLP:journals/corr/abs-2102-04664},  for two  main reasons:
\begin{enumerate}
\item \cxglue is carefully de-duplicated~\cite{shi2021neural}. Prior datasets like TL-CodeSum~\cite{hu2018summarizing} have duplicates~\cite{shi2021neural} in training, testing, and validation partitions. Duplication can inflate measured performance~\cite{shi2021neural, allamanis2019adverse}. 
\item We need a multilingual dataset to prove the effectiveness of multilingual fine-tuning. None of the existing datasets~\cite{hu2018summarizing,leclair2019neural} is multilingual. 
\end{enumerate}
Table~\ref{dataset} presents the number of training, testing and validation instances for each language.
in \cxglue.

\begin{table}[h]

\centering
\resizebox{.8\columnwidth}{!}{%
\renewcommand{\arraystretch}{1.2}

\begin{tabular}{lllcc}

\hline
\multicolumn{1}{c}{\begin{tabular}[c]{@{}c@{}}Programming \\ language\end{tabular}} & \multicolumn{1}{c}{Training} & \multicolumn{1}{c}{Dev}   & \multicolumn{1}{c}{Test} & \multicolumn{1}{c}{\begin{tabular}[c]{@{}c@{}}Candidate \\ codes*\end{tabular}}  \\ \hline

Ruby                                                                            & 24,927                        & 1,400                      & 1,261 &     4,360                 \\ 
JavaScript                                                                            & 58,025                        & 3,885                      & 3,291   &   13,981                \\ 
Java                                                                                  & 164,923                       & 5,183                      & 10,955   &   40,347               \\
Go                                                                                    & 167,288                       & 7,325                      & 8,122   &     28,120              \\
PHP                                                                                   & 241,241                       & 12,982                     & 14,014  &    52,660               \\
Python                                                                                & 251,820                       & 13,914                     & 14,918   &    43,827              \\ \hline
\multicolumn{4}{l}{*Candidate codes are only used for code retrieval task}
\end{tabular}
}
\vspace{0.05in}
\caption{{\em {\small \cxglue dataset}}}
\label{dataset}
\vspace{-0.2in}
\end{table}

\noindent{\underline{\em Model \& Fine-tuning}}
Feng \etal use a transformer-based encode-decoder architecture for  the code summarization task~\cite{feng2020codebert}. 
The encoder is all ready well-trained in the pre-training stage; 
for  fine-tuning, the encoder is primed with weights from pre-training. Now, 
the transformer model is given the input code token sequence and asked to generate the comment, as in the Neural Machine Translation (NMT) problem. 
We fine-tune using the \cxglue paired samples. 
During fine-tuning,  the decoder is trained auto-regressively,  using next-token cross-entropy loss. 
Feng \etal use smooth BLEU-4~\cite{lin2004orange} for the evaluations of the models.
Subsequently, We replace the pre-trained \cbert with pre-trained \gcbert in the encoder while evaluating the effectiveness of multilingual fine-tuning with \gcbert.



%

\noindent{\underline{\em Why baseline with \cbert for code summarization?}} Feng \etal compare \cbert with other popular encoder-decoder based (\eg LSTM~\cite{sutskever2014sequence}, Transformer~\cite{vaswani2017attention}, RoBERTa~\cite{liu2019roberta}) models; \cbert handily
beats all of them~\cite{feng2020codebert}. Thus, \cbert  is a good baseline to measure  the value of multilingual finetuning. 
\cbert also does very well on prior datasets: using smoothed 
Sentence BLEU-4, we found that \cbert reaches 44.89 on TL-Codesum~\cite{hu2018summarizing}, and 32.92 on Funcom~\cite{leclair2019neural}\footnote{As reported in~\cite{gros2020code,shi2021neural}, 
measurement approaches vary across papers, and these numbers may differ from prior results: we use smoothed sentence BLEU-4 everywhere in our paper.}.
TL-Codesum has high degree of duplicates; we found that Funcom also does, but just in the comments. 
\cxglue has very little duplication, which makes it more challenging, and also more reliable. Note that \gcbert \emph{does not report} any performance on the code summarization task, and so we had to measure it. 



\subsection{Code Search}

%

\noindent{\underline {\em The Task}} Given a natural language query, find the  semantically closest code sample from a large set of candidates. 
Vector-based information retrieval methods can be used
here along with BERT-style encoders. \cbert was shown to perform quite well; the best published performance is reported
by \gcbert~\cite{guo2020graphcodebert}  (\cbert augmented with graph representations). 
We study the value of multilingual fine-tuning for  both \cbert and \gcbert (pre-training of both models was discussed earlier in Section~\ref{cgcmodels}).

\noindent{\underline{\em The Dataset: }}
Guo \etal adapt the same CodeSearchNet~\cite{husain2019codesearchnet} dataset, with some additional data for candidate codes~\cite{guo2020graphcodebert}. Note that it is basically the same dataset we used for code summarization except the candidate codes. 


 


\noindent{\underline{\em Model \& Fine-tuning}} 
We use Guo \etal's \gcbert model, which  at the time of submission is the best performing  model with code and parameters available, and so is fine-tunable.  
The fine-tuning data is code (PL)  matched with (NL) comments, from \cxglue. The pre-trained \gcbert embedding vector is calculated
for each PL and NL part. 
During fine-tuning, Guo \etal take a minibatch of (say $n$) NL query vector, along with $n$ (correct answers) PL answer vectors. $n^2$ dot products
are calculated; the
embedding vectors are then full-stack trained to give "1" normalized dot product for the matches, and "0" for the mis-matches. 
For the actual retrieval, \gcbert  calculates the vector embedding of a given query, and
simply retrieves candidates ranked by the  dot-product distance from the query vector.  



\subsection{Method Name Prediction}

\noindent{\underline {\em The Task}} as introduced by Allamanis \etal ~\cite{allamanis2016convolutional} as the ``extreme summarization'' problem,
the task is to predict the function name given the body. 

\noindent{\underline{\em The Dataset: }} We adapt the \cxglue dataset by extracting the function name and asking the model to find the name given the function body. 
Following~\cite{allamanis2016convolutional}, the function names are broken into subtokens using BPE~\cite{sennrich2015neural} (we've used
BPE tokenization for all tasks).  This problem then becomes very similar to code summarization.  

\noindent{\underline{\em Model \& Fine-tuning}} 
Previously Code2Seq~\cite{alon2018codeseq} and Code2Vec~\cite{alon2019code2vec} have worked on this problem. All prior works~\cite{allamanis2016convolutional,alon2018codeseq,alon2019code2vec} use a mono-lingual datasets, which are not suitable for our experiment. 
We use the same model we used for summarization, except we now 
learn to sequentially generate the method name, subtoken by  subtoken. 
We use F1-score for the evaluation. For example, the function name ``createLocal'' is broken into two sub tokens (\ie create and Local), and the model predicts only ``create''. Hence, the precision, recall, and F1-score are 1.0, 0.5, and 0.66, respectively.

%% file: result.tex
\section{Results}
In this section, we evaluate multilingual fine-tuning  for the baselines for the tasks enumerated above. 

\subsection{Code Summarization}
\label{csumr}

We apply multilingual fine-tuning on the \cxglue dataset. We first \emph{replicate} the summarization task by (monolingually) 
fine-tuning the available pre-trained \cbert 
model  for six languages\footnote{We use the publicly available \cbert implementation and dataset, \url{https://github.com/microsoft/CodeXGLUE/tree/main/Code-Text/code-to-text}}. 
We replicate the fine-tuning stage for 2 reasons: 

\begin{enumerate}
\item We want to account for any hardware or environmental bias (\eg we have a different set of GPUs than the original paper. We fine-tune with NVIDIA TITAN RTX, while Feng \etal~\cite{feng2020codebert} use NVIDIA Tesla V100). 
\item We use a pairwise two-sample statistical test (as described in~\cite{roy2021reassessing}, it is more precise than just comparing test-set summary statistics) to gauge differences. 
This requires a performance measurement for each test sample, which the repository did not include.
\end{enumerate}  
Our BLEU-4 numbers  for monolingual training
were close to reported numbers, with some differences; but we do obtain the same overall score (17.83) (\cref{tbl:csum}, leftmost 2 columns). 

We use the same, per-language test sets to compare  monolingual and multilingual fine-tuning. 
The validation set, however, is a single multilingual one combining all the monolingual validation sets. 
\Cref{tbl:csum} shows that multilingual fine-tuning improves performance, even for high-resource languages (with more than 100K training instances). 
With \cbert, multilingual fine-tuning gains 2.5\%-17.5\% over monolingual fine-tuning, 
 for all languages,  yielding a  6.90\% overall improvement (4.48\% weighted improvement)\footnote{The \cbert paper simply averages the BLEU across languages to report the ``overall'' number; our \emph{weighted average}  weights each BLEU by the number of samples in that language.}. 
 With the more advanced \gcbert, we  see  smaller gains, although the relative gains span a wide range.

\begin{table*}[h]

\centering

\resizebox{\textwidth}{!}{%
\renewcommand{\arraystretch}{1.2}

\begin{tabular}{lccccccccccc}

\hline
\multicolumn{1}{c}{Language}                                                            & \multicolumn{1}{c}{\begin{tabular}[c]{@{}c@{}}\cbert 			\\ (reported)\end{tabular}} & \multicolumn{1}{c}{\begin{tabular}[c]{@{}c@{}}\cbert 			\\ (re-trained)\end{tabular}} & \multicolumn{1}{c}{\mlcbert } & \multicolumn{1}{c}{Improvement} & \multicolumn{1}{c}{\begin{tabular}[c]{@{}c@{}}Effect 			\\ Size\end{tabular}} & \multicolumn{1}{c}{\begin{tabular}[c]{@{}c@{}}p-value 			\\ (adjusted)\end{tabular}} & \multicolumn{1}{c}{\gcbert} & \multicolumn{1}{c}{\mlgcbert} & \multicolumn{1}{c}{Improvement} & \multicolumn{1}{c}{\begin{tabular}[c]{@{}c@{}}Effect 			\\ Size\end{tabular}} & \multicolumn{1}{c}{\begin{tabular}[c]{@{}c@{}}p-value 			\\ (adjusted)\end{tabular}} \\ \hline
Ruby                                                                                      & 12.16                                                                                  & 12.53                                                                                    & 14.75                                  & +17.72\%                         & 0.055 & \textless{}0.001                                                                      & 12.62                       & \textbf{14.95}                                     & +18.46\%                         & 0.055                                                                          & \textless{}0.001                                                                      \\
JS                                                                                        & 14.90                                                                                  & 13.86                                                                                    & \textbf{15.80                                 } & +14.00\%                         & 0.016 & \textless{}0.001                                                                      & 14.79                       & 15.79                                     & +6.76\%                          & 0.016                                                                          & 0.014                                                                                 \\
Java                                                                                      & 17.65                                                                                  & 18.72                                                                                    & \textbf{20.11                                 } & +7.43\%                          & 0.016 & \textless{}0.001                                                                      & 19.22                       & 19.91                                     & +3.59\%                          & 0.016                                                                          & \textless{}0.001                                                                      \\
Go                                                                                        & 18.07                                                                                  & 18.15                                                                                    & 18.77                                  & +3.42\%                          & 0.010 & \textless{}0.001                                                                      & 18.40                       & \textbf{18.92 }                                    & +2.83\%                          & 0.010                                                                          & \textless{}0.001                                                                      \\
PHP                                                                                       & 25.16                                                                                  & 25.48                                                                                    & \textbf{26.23                                 } & +2.94\%                          & 0.012 & \textless{}0.001                                                                      & 25.45                       & 26.15                                     & +2.75\%                          & 0.012                                                                          & \textless{}0.001                                                                      \\
Python                                                                                    & \textbf{19.06                                                                                 } & 18.25                                                                                    & 18.71                                  & +2.52\%                          & 0.022 & \textless{}0.001                                                                      & 18.02                       & 18.90                                     & +4.88\%                          & 0.022                                                                          & \textless{}0.001                                                                      \\ \hline
\multicolumn{1}{l}{Overall}                                                      & \multicolumn{1}{c}{17.83}                                                             & \multicolumn{1}{c}{17.83}                                                               & \multicolumn{1}{c}{19.06}             & \multicolumn{1}{c}{+6.90\%}     & \multicolumn{1}{c}{\multirow{2}{*}{0.016}}                                & \multicolumn{1}{c}{\multirow{2}{*}{\textless{}0.001}}                                & \multicolumn{1}{c}{18.08}  & \multicolumn{1}{c}{\textbf{19.10}}                & \multicolumn{1}{c}{+5.64\%}     & \multicolumn{1}{c}{\multirow{2}{*}{0.016}}                                    & \multicolumn{1}{c}{\multirow{2}{*}{\textless{}0.001}}                                \\ 
\multicolumn{1}{l}{\begin{tabular}[c]{@{}l@{}}Overall\\ (weighted)\end{tabular}} & \multicolumn{1}{l}{\begin{tabular}[c]{@{}l@{}}Not\\ Reported\end{tabular}}            & \multicolumn{1}{c}{19.85}                                                               & \multicolumn{1}{c}{20.74}             & \multicolumn{1}{c}{+4.48\%}     & \multicolumn{1}{c}{}                                                          & \multicolumn{1}{c}{}                                                                 & \multicolumn{1}{c}{19.98}  & \multicolumn{1}{c}{\textbf{20.76}}                & \multicolumn{1}{c}{+3.90\%}     & \multicolumn{1}{l}{}                                                          & \multicolumn{1}{c}{}                                                                 \\ \hline
\multicolumn{12}{l}{*Evaluation criteria followed by \cxglue~\cite{DBLP:journals/corr/abs-2102-04664} and \cbert~\citep{feng2020codebert}}  
\end{tabular}

}
\vspace{0.05in}
\caption{{\em {\small Effectiveness of multi-lingual fine-tuning for code summarization task. Note that p-values are B-H corrected}}}
\vspace{-0.2in}
\label{tbl:csum}
\end{table*}


We use a one-sided (AH: monolingual < multilingual) \underline{pairwise} Wilcoxon signed-rank test
(thus avoiding the \underline{corpus-level} measurement pitfalls noted in~\cite{roy2021reassessing}). 
Null hypothesis is rejected for all six languages, for \cbert. For \gcbert, it's rejected overall, and for every language;
except for Javascript, where the p-value is 0.014 (all after B-H correction). 

Thus our measurement indicates that multilingual fine-tuning 
provides a statistically significant improvement over monolingual training. We find rather low  effect sizes using Cliff's Delta~\cite{macbeth2011cliff}. 
While we report the effect size for the sake of completeness, this is not a major concern: 
we note that \emph{all gains are statistically highly significant}. 
We also emphasize that even the minor 
improvements provided here by multilingual training (which is broadly compatible with a range of settings) constitute a relevant and potentially widely useful result. 
Roy \emph{et al}~\cite{roy2021reassessing} have previously noted that small gains in BLEU-4 may not be perceptible to humans as
increased text quality; nevertheless, we note that  natural language translation (which is now widely used) attained
 high performance levels based on decades of incremental progress; this result and others below
provide evidence that  multilingual training could be an important step in the progress towards more useful automated
tools.  Finally, we note 
 that BLEU-4 gains are higher for low-resource language (\eg 17.7\% for Ruby), and lower for high-resource languages (\eg 2.5\% for Python), as expected. 


\noindent{\underline{\em Comparing multi-lingual \cbert with other models}}
Code summarization is widely studied---there are many models for this task; our specific focus 
here is to understand if multilingual fine-tuning provides benefits, using a high-quality token-sequence model and dataset. 
So we focus comparisons on the papers which report performance on \cxglue dataset, and use a token-sequence inductive bias:
comparing against all models is beyond the scope of this paper. 
We compare multi-lingual \cbert (\mlcbert) and \gcbert (\mlgcbert) with other models that  have been published in peer-reviewed venues; 
among them, four apply pre-training strategies~\cite{liu2019roberta,feng2020codebert,ahmad-etal-2021-unified,qi2021prophetnet}. 
We achieve the best overall performance (table~\ref{comp1}), outperforming all the models, and  for four specific languages (\ie Ruby, Java, Go and PHP).

There is one other system, CoTexT~\cite{phan2021cotext}  which claims (in an unpublished, non-peer-reviewed report) 
 better performance than us for just Python~\cite{phan2021cotext}, but is worse overall. We will include it for
comparison once it is published in a peer-reviewed venue.

This table also provides evidence supporting the effectiveness of multilingual fine-tuning.


\begin{table}[b]

\centering
\resizebox{\columnwidth}{!}{%
\renewcommand{\arraystretch}{1.2}
\begin{tabular}{lccccccc}
\hline
\multicolumn{1}{c}{Models}  & \multicolumn{1}{c}{Overall} & \multicolumn{1}{c}{Ruby} & \multicolumn{1}{c}{JavaScript}    & \multicolumn{1}{c}{Go}    & \multicolumn{1}{c}{Python} & \multicolumn{1}{c}{Java}  & \multicolumn{1}{c}{PHP}   \\ \hline

\mlgcbert           & \textbf{19.10}                        & \textbf{14.95}                     & 15.79 & \textbf{18.92 }                     & 18.90                       & 19.91 & 26.15\\

\mlcbert           & 19.06                        & 14.75                     & 15.80                     & 18.77                      & 18.71                       & \textbf{20.11                     } & \textbf{26.23}                      \\
ProphetNet-X~\cite{qi2021prophetnet}                  & 18.54                        & 14.37                     & \textbf{16.60                      } & 18.43                      & 17.87                       & 19.39                      & 24.57                      \\
PLBART~\cite{ahmad-etal-2021-unified}                        & 18.32                        & 14.11                     & 15.56                      & 18.91                      & \textbf{19.30}                       & 18.45                      & 23.58                      \\

\gcbert~\cite{guo2020graphcodebert}                        & 18.08                        & 12.62 & 14.79 & 18.40                      & 18.02                       & 19.22                      & 25.45\\

\cbert~\cite{feng2020codebert}                         & 17.83                        & 12.16                     & 14.90                      & 18.07                      & 19.06                       & 17.65                      & 25.16                      \\
RoBERTa~\cite{liu2019roberta}                       & 16.57                        & 11.17                     & 11.90                      & 17.72                      & 18.14                       & 16.47                      & 24.02                      \\
Transformer~\cite{vaswani2017attention}                   & 15.56                        & 11.18                     & 11.59                      & 16.38                      & 15.81                       & 16.26                      & 22.12                      \\ 
\multicolumn{1}{l}{Seq2Seq~\cite{sutskever2014sequence}} & \multicolumn{1}{c}{14.32}   & \multicolumn{1}{c}{9.64} & \multicolumn{1}{c}{10.21} & \multicolumn{1}{c}{13.98} & \multicolumn{1}{c}{15.93}  & \multicolumn{1}{c}{15.09} & \multicolumn{1}{c}{21.08} \\ \hline
\end{tabular}
}
\vspace{0.05in}

\caption{{\em {\small Comparison to existing models, on \cxglue dataset}}}
\vspace{-0.2in}
\label{comp1}
\end{table}

\subsection{Code Search}
We study the gains from multilingual fine-tuning using two  pre-trained models (\ie \cbert \& \gcbert). We multilingually fine-tune both models using the publicly available code \& dataset~\footnote{https://github.com/microsoft/CodeBERT/tree/master/GraphCodeBERT/codesearch}. As we did for code summarization, we re-trained the baseline models, to get performance numbers
for each case in the test set (to enable pairwise two-sample testing).
We use the same test sets for both monolingual and multilingual training to evaluate our approach. During the training, \gcbert uses a matrix of dimension $|query|*|candidate\_codes|$. We could not use the full merged validation set (as we did for the code summarization task) because that makes the query and 
candidate code sets too large; the resulting matrix could not fit on our GPU server.
We used a down-sampled validation set comprising six monolingual validation sets with 10K query and 50K candidate codes each. However, we did not face any issue while testing because we did not merge the test sets. 

We report both the published values, and our replication; we need the replication to measure pairwise  gains. 
Though \cbert and \gcbert both work on sequence of code tokens, \gcbert creates a rudimentary data-flow graph, once it's told the programming language. 

Table~\ref{csearch} shows that multilingual fine-tuning improves the mean reciprocal rank for all languages except Go with \cbert. The improvement for Ruby, JavaScript, and Java are statistically significant. We found similar results for \gcbert exhibiting improvement for Ruby, JavaScript, Java, and Python; but with \gcbert both Go and PHP showed performance declines. 
However, overall, both showed statistically signficant improvements (p < 0.001); but the 
 improvement for \gcbert(1.54\%) is lower than \cbert(2.74\%). 
Finally, we note that our numbers for \cbert differ from the performance reported for  on the \cxglue leaderboard. 
This is because \cxglue benchmark uses only Python, and is based on
a restricted setting where identifier names are left out. 
\cxglue team argues that this abstraction enables them to stress-test the generalization ability of a model; however,
here we consider an unmodified setting where someone gives an natural language query and wishes to find
``natural'' code with variable names intact.


\begin{table*}[h]

\centering
\resizebox{\textwidth}{!}{%
\renewcommand{\arraystretch}{1.2}

\begin{tabular}{lcccccccccccc}
\hline
\multicolumn{1}{c}{Language}                                                      & \multicolumn{1}{c}{\begin{tabular}[c]{@{}c@{}}\cbert\\ (published)~\cite{guo2020graphcodebert}\end{tabular}} & \multicolumn{1}{c}{\begin{tabular}[c]{@{}c@{}}\cbert \\ (re-trained)\end{tabular}} & \multicolumn{1}{c}{\begin{tabular}[c]{@{}c@{}}\mlcbert\end{tabular}} & \multicolumn{1}{c}{Improvement} & \multicolumn{1}{c}{\begin{tabular}[c]{@{}c@{}}Effect \\ Size\end{tabular}}                          & \multicolumn{1}{c}{\begin{tabular}[c]{@{}c@{}}p-value\\ (adjusted)\end{tabular}} & \multicolumn{1}{c}{\begin{tabular}[c]{@{}c@{}}\gcbert\\ (published)~\cite{guo2020graphcodebert}\end{tabular}} & \multicolumn{1}{c}{\begin{tabular}[c]{@{}c@{}}\gcbert\\ (re-trained)\end{tabular}} & \multicolumn{1}{c}{\begin{tabular}[c]{@{}c@{}}\mlgcbert\end{tabular}} & \multicolumn{1}{c}{Improvement} & \multicolumn{1}{c}{\begin{tabular}[c]{@{}c@{}}Effect \\ Size\end{tabular}}                         & \multicolumn{1}{c}{\begin{tabular}[c]{@{}c@{}}p-value\\ (adjusted)\end{tabular}} \\ \hline
Ruby                                                                                & 0.679                                                                           & 0.677                                                                              & 0.732                                                                         & +8.12\%                          & 0.072                                           & \textless{}0.001                                                                  & 0.703                                                                            & 0.708                                                                              & \textbf{0.738 }                                                                         & +4.24\%                          & 0.039                                           & \textless{}0.001                                                                  \\
JavaScript                                                                          & 0.620                                                                           & 0.616                                                                              & 0.643                                                                         & +4.38\%                          &  0.034                                          & \textless{}0.001                                                                  & 0.644                                                                            & 0.644                                                                              & \textbf{0.660}                                                                          & +2.48\%                          &  0.019                                          & 0.004 \\
Java                                                                                & 0.676                                                                           & 0.676                                                                              & 0.697                                                                         & +3.11\%                          & 0.026                                         & \textless{}0.001                                                                  & 0.691                                                                            & 0.693                                                                              & \textbf{0.710 }                                                                         & +2.45\%                          &  0.022                                        & \textless{}0.001                                                                  \\
Go                                                                                  & 0.882                                                                           & 0.885                                                                              & 0.885                                                                         & 0\%                              & -0.003                                          & 0.550 & \textbf{0.897                                                                           } & 0.894                                                                              & 0.894                                                                          & 0\%                              & -0.002                                          & 0.724                                                                           \\
PHP                                                                                 & 0.628                                                                           & 0.629                                                                              & 0.635                                                                         & +0.95\%                          & 0.009                                     & 0.003 & \textbf{0.649                                                                           } & 0.648                                                                              & 0.646                                                                          & -0.31\%                          &  -0.002                                        & 0.904 \\
Python                                                                              & 0.672                                                                           & 0.676                                                                              & 0.678                                                                         & +0.30\%                          &  0.004                                        & 0.050 & 0.692                                                                            & 0.692                                                                              & \textbf{0.695                                                                         } & +0.43\%                          & 0.005                                       & 0.300 \\ \hline
\multicolumn{1}{l}{Overall*}                                                      & \multicolumn{1}{c}{0.693}                                                      & \multicolumn{1}{c}{0.693}                                                         & \multicolumn{1}{c}{0.712}                                                    & \multicolumn{1}{c}{+2.74\%}     & \multicolumn{1}{c}{\multirow{2}{*}{ 0.013}} & \multicolumn{1}{c}{\multirow{2}{*}{\textless{}0.001}}                            & \multicolumn{1}{c}{0.713}                                                       & \multicolumn{1}{c}{0.713}                                                         & \multicolumn{1}{c}{\textbf{0.724}}                                                     & \multicolumn{1}{c}{+1.54\%}     & \multicolumn{1}{c}{\multirow{2}{*}{0.007}} & \multicolumn{1}{c}{\multirow{2}{*}{\textless{}0.001}}                            \\ 
\multicolumn{1}{l}{\begin{tabular}[c]{@{}l@{}}Overall \\ (weighted)\end{tabular}} & \multicolumn{1}{c}{\begin{tabular}[c]{@{}l@{}}Not \\ Reported\end{tabular}}    & \multicolumn{1}{c}{0.692}                                                         & \multicolumn{1}{c}{0.702}                                                    & \multicolumn{1}{c}{+1.42\%}     & \multicolumn{1}{c}{}                           & \multicolumn{1}{c}{}                                                             & \multicolumn{1}{c}{\begin{tabular}[c]{@{}l@{}}Not \\ Reported\end{tabular}}     & \multicolumn{1}{c}{0.709}                                                         & \multicolumn{1}{c}{\textbf{0.715}}                                                     & \multicolumn{1}{c}{+0.80\%}     & \multicolumn{1}{c}{}                           & \multicolumn{1}{c}{}                                                             \\ \hline
\multicolumn{13}{l}{*Evaluation criteria followed by \gcbert~\cite{guo2020graphcodebert}}                                                                                                                                                                                                                                                                                                                                                                                                                                                                                                                                                                                                                                                                                                                                                                                                                                                                                                
\end{tabular}
}
\vspace{0.05in}
\caption{{\em {\small Effectiveness of multi-lingual fine-tuning for code search task. Note that p-values are BH-corrected}}}

\label{csearch}
\vspace{-0.2in}
\end{table*}


\subsection{Method Name Prediction}

As for the previous two tasks, we try multilingual fine-tuning for method name prediction for \cbert.   
Here, too, we find evidence supporting the conclusion that multilingual training provides improvement for all the languages (Table~\ref{mpred}). Non-parametric pairwise 
improvements are significant for Ruby, JavaScript, and Java. We also  note 
observe relatively greater effect size for Ruby and JavaScript. 
Note that we achieve highest improvement for JavaScript because many functions therein are anonymous lambdas, 
since these functions have no names, they are not useful, and this diminishes available the JavaScript training set  relative to other tasks 
(lambdas still have summaries, and can be used for other tasks). Therefore, multilingual fine-tuning increases the dataset diversity and boosts JavaScript method name prediction performance.

\begin{table}[h]
\centering
\resizebox{\columnwidth}{!}{%
\renewcommand{\arraystretch}{1.2}

\begin{tabular}{lccccccccc}
\hline
\multicolumn{1}{c}{\multirow{2}{*}{Language}}                                    & \multicolumn{3}{c}{\cbert}                                                               & \multicolumn{3}{c}{\mlcbert}                                                      & \multicolumn{1}{c}{\multirow{2}{*}{\begin{tabular}[c]{@{}c@{}}F-Score \\ Improvement\end{tabular}}} & \multicolumn{1}{c}{\multirow{2}{*}{\begin{tabular}[c]{@{}c@{}}Effect \\ Size\end{tabular}} }         & \multicolumn{1}{c}{\multirow{2}{*}{\begin{tabular}[c]{@{}c@{}}p-value \\ (adjusted)\end{tabular}}} \\ 
\multicolumn{1}{c}{}                                                             & \multicolumn{1}{c}{Precision} & \multicolumn{1}{c}{Recall} & \multicolumn{1}{c}{F-Score} & \multicolumn{1}{c}{Precision} & \multicolumn{1}{c}{Recall} & \multicolumn{1}{c}{F-Score} & \multicolumn{1}{c}{}                                                                                & \multicolumn{1}{c}{}                           & \multicolumn{1}{c}{}                                                                               \\ \hline
Ruby                                                                               & 0.44                           & 0.40                       & 0.41                        & 0.53                          & 0.49                       & \textbf{0.49 }                       & 20.59\%                                                                                              & 0.112                                          & \textless{}0.001                                                                                    \\
JavaScript                                                                         & 0.30                          & 0.24                       & 0.26                        & 0.45                          & 0.40                       & \textbf{0.41}                        & 59.00\%                                                                                              & 0.215                                          & \textless{}0.001                                                                                    \\
Java                                                                               & 0.54                          & 0.51                       & 0.51                        & 0.56                          & 0.52                       & \textbf{0.52}                        & 2.22\%                                                                                               & 0.016                                         & \textless{}0.001                                                                                    \\
Go                                                                                 & 0.54                          & 0.52                       & 0.52                        & 0.56                          & 0.53                       & 0.52                        & 1.67\%                                                                                               &  0.015                                         & 0.004 \\
PHP                                                                                & 0.56                          & 0.53                       & 0.52                        & 0.57                          & 0.53                       & \textbf{0.53 }                       & 1.30\%                                                                                               & 0.009                                        & 0.004 \\
Python                                                                             & 0.49                          & 0.45                       & 0.45                        & 0.50                          & 0.45                       & \textbf{0.46 }                       & 1.60\%                                                                                               & 0.011                                         & 0.002 \\ \hline
\multicolumn{1}{l}{Overall}                                                      & \multicolumn{1}{c}{0. 48}     & \multicolumn{1}{c}{0.44}  & \multicolumn{1}{c}{0.44}   & \multicolumn{1}{c}{0.53}     & \multicolumn{1}{c}{0.49}  & \multicolumn{1}{c}{\textbf{0.49}}   & \multicolumn{1}{c}{10.09\%}                                                                         & \multicolumn{1}{c}{\multirow{2}{*}{0.024}} & \multicolumn{1}{c}{\multirow{2}{*}{\textless{}0.001}}                                              \\ 
\multicolumn{1}{l}{\begin{tabular}[c]{@{}l@{}}Overall\\ (weighted)\end{tabular}} & \multicolumn{1}{c}{0. 52}     & \multicolumn{1}{c}{0.48}  & \multicolumn{1}{c}{0.48}   & \multicolumn{1}{c}{0.54}     & \multicolumn{1}{c}{0.50}  & \multicolumn{1}{c}{\textbf{0.50}}   & \multicolumn{1}{c}{3.37\%}                                                                          & \multicolumn{1}{c}{}                           & \multicolumn{1}{c}{}                                                                               \\ \hline
\end{tabular}
}
\vspace{0.05in}
\caption{{\em {\small Effectiveness of multi-lingual fine-tuning for method naming task. Note that p-values are adjusted using Benjamini-Hochberg}}}
\label{mpred}
\vspace{-0.2in}
\end{table}


\begin{table}[h]

\centering
\resizebox{\columnwidth}{!}{%
\renewcommand{\arraystretch}{1.2}

\begin{tabular}{lll}
\hline

\multicolumn{3}{l}{\begin{tabular}[c]{@{}l@{}}
\textbf{Example:1}\\
\textit{//set the values from an Array}\\

public void setValues* ( Array arr ) \{\\
\hspace{.5cm}	\textit{//we omit intermediate lines to fit in the paper}\\
\hspace{.5cm}	\textit{//original code \href{https://github.com/Unidata/thredds/blob/d2d68f9eee87f345625211324d71d5dc3e162ee1/cdm/src/main/java/ucar/nc2/Attribute.java\#L548-L596
}{\link{here}}}\\
\}

\end{tabular}} \\ 

\hline
\multicolumn{3}{c}{\emph{Code Summarization}}                                                                                                                                                            \\
\multicolumn{2}{l}{Models \& comments}                                                                                                    & BLEU-4                                                \\ \hline
\multicolumn{2}{l}{Gold: set the values from an Array}                                                      & NA                                                    \\
\multicolumn{2}{l}{\cbert: Sets the values of the array .}                                                                            & 25                                                  \\
\multicolumn{2}{l}{\mlcbert: Set the values from an array .}                                                                      & 84                                                \\ \hline
\multicolumn{3}{c}{\emph{Code Search}}                                                                                                                                                                 \\ 
\multicolumn{2}{l}{Models}                                                                                                                & MRR                                                   \\ \hline
\multicolumn{2}{l}{\gcbert}                                                                                                                 & 0.33                                                  \\
\multicolumn{2}{l}{\mlgcbert}                                                                                                   & 1.00                                                  \\ \hline
\multicolumn{3}{c}{\emph{Method Name Prediction}}                                                                                                                                                      \\ 
\multicolumn{1}{l}{Models \& method name}                                   & \multicolumn{1}{l}{Sub tokens}                           & \multicolumn{1}{l}{F-Score}                          \\ \hline
Gold: setValues                                                                 & set Values                                                 & NA                                                    \\
\cbert: setArrayValue                                                                  & set Array Value                                                     & 0.40                                                  \\ 
\multicolumn{1}{l}{\mlcbert: setValue}                           & \multicolumn{1}{l}{set Value}                           & \multicolumn{1}{l}{0.50}                             \\ \hline

\multicolumn{3}{l}{\begin{tabular}[c]{@{}l@{}}
\textbf{Example:2}\\

\textit{//Registers set injection point .}\\
public void registerPetiteSetInjectionPoint* ( final String beanName, final String property ) \{\\
\hspace{.5cm}	\textit{//we omit intermediate lines to fit in the paper}\\
\hspace{.5cm}	\textit{//original code \href{https://github.com/oblac/jodd/blob/85ad7f813ec0e07ecd27042aeb47ff2047631fa5/jodd-petite/src/main/java/jodd/petite/PetiteBeans.java\#L585-L598
}{\link{here}}}\\
\}

\end{tabular}} \\ 

\hline
\multicolumn{3}{c}{\emph{Code Summarization}}                                                                                                                                                            \\
\multicolumn{2}{l}{Models \& comments}                                                                                                    & BLEU-4                                                \\ \hline
\multicolumn{2}{l}{Gold: Registers set injection point .}                                                      & NA                                                    \\
\multicolumn{2}{l}{\cbert: Register a set of set InjectionPoint .}                                                                            & 19                                                 \\
\multicolumn{2}{l}{\mlcbert: Register a set injection point .}                                                                      & 60                                                  \\ \hline
\multicolumn{3}{c}{\emph{Code Search}}                                                                                                                                                                 \\ 
\multicolumn{2}{l}{Models}                                                                                                                & MRR                                                   \\ \hline
\multicolumn{2}{l}{\gcbert}                                                                                                                 & 0.50                                                  \\
\multicolumn{2}{l}{\mlgcbert}                                                                                                   & 1.00                                                  \\ \hline
\multicolumn{3}{c}{\emph{Method Name Prediction}}                                                                                                                                                      \\ 
\multicolumn{1}{l}{Models \& method name}                                   & \multicolumn{1}{l}{Sub tokens}                           & \multicolumn{1}{l}{F-Score}                          \\ \hline
Gold: registerPetiteSetInjectionPoint                                                                 & register Pet ite Set In jection Point                                                 & NA                                                    \\
\cbert: addPropertyInjectionPoint                                                                  & add Property In jection Point                                                     & 0.50 \\ 
\multicolumn{1}{l}{\mlcbert: setPropertyInjectionPoint}                           & \multicolumn{1}{l}{set Property In jection Point}                           & \multicolumn{1}{l}{0.57}                             \\ \hline

\multicolumn{3}{l}{\emph{*``registerPetiteSetInjectionPoint'' \& ``setValues'' tokens are abstracted for method name prediction task}}

\end{tabular}
}

\vspace{0.05in}
\caption{{\em {\small Examples exhibiting the effectiveness of multilingual training}}}
\label{exmp}
\vspace{-0.2in}

\end{table}


\subsection{Two Illustrative Examples}
We used the same dataset for all tasks; for illustration, we show 
(Table~\ref{exmp})  two test instances where all the tasks show improved performance  from multilingual fine-tuning. In code summarization task, the monolingual fine-tuning scores 25 BLEU-4 in Example 1. \cbert produces a semantically wrong comment where multilingual fine-tuning generates the semantically correct solution. Note that the BLEU-4 is 84 for the second example because of the missing period in the gold standard (BLEU-4 is case-insensitive). Multilingual fine-tuning also helps the code search problem by increasing the MRR from 0.33 (Rank:3) to 1.00 (Rank:1). We also observe performance improvement from the method name prediction task. The gold standard consists of two sub tokens (\ie set and Values), and mono-lingual fine-tuning generates three (\ie set, Array, and Value), one of them is exact match. On the other hand, multilingual fine-tuning removes the extra ``Array'' subtoken and produces two subtokens(\ie set and Value) resulting in the F-score 0.50. We observe a similar result in example 2. Note that like BLEU-4, our method name prediction metric is also case-insensitive.



\takeaway{3}{Multilingual fine-tuning is likely to increase diversity and help the models perform better than those trained with smaller mono-lingual datasets, especially for low-resource languages, irrespective of the task.}

%% file: interpret.tex
\section{Interpreting results, and Threats}
In this section we consider several 
issues that are relevant to the observed performance of 
multilingual training, such as model choice, dataset duplication, performance metrics, generalization, and different training strategies for the models.

\subsection{Does multilingual fine-tuning help with other models?}

There are several models, including CoTexT~\cite{phan2021cotext}, ProphetNet-X~\cite{qi2021prophetnet}, and PLBART~\cite{ahmad-etal-2021-unified} which
report higher performance than \cbert~\cite{feng2020codebert} model for the code summarization task. The models  for all these tasks were fine-tuned using 
monolingual datasets, so we might expect that multilingual fine-tuning should improve performance. These experiments would require a substantial investment
of compute energy and is left for future work. We focused on \cbert (and also \gcbert on some tasks). 
We did some preliminary experiments with multilingual fine-tuning on PLBART. In our preliminary study, did see the same gains for low-resource language  (Ruby, 5\% gain). 
However, we found a 0.55\% overall loss, which is inconsistent with what we observe with \mlcbert (6.90\% overall improvement) \& \mlgcbert (5.64\% overall improvement). More study is needed. 

 
\takeaway{4}{Multilingual fine-tuning could benefit a broad range of models. We find gains for \cbert and \gcbert, but more data is
required for other models.}


\subsection{Threats: Risk of data duplication?}
Data duplication can lead to poor-quality estimates of performance, especially when data is duplicated
across training \& test; even duplication just within test data risks higher variance in the
 estimates.  Allamanis finds that performance metrics are highly inflated when test data has duplicates, and advocates de-duplicating datasets, for more robust results~\cite{allamanis2019adverse}. Shi \etal also discusses the impact of duplication in code summarization task~\cite{shi2021neural}.

Sadly, there is a large amount of copied code on GitHUB~\cite{lopes2017dejavu};  inattentively combining different datasets 
harvested from GitHUB can lead to undesirable levels of duplication in the merged dataset. 
Fortnuately, \cxglue is atually a \emph{carefully de-duplicated} dataset; performance estimates therein are thus
more robust. 
Combining \emph{multilingual} data  is unlikely to
introduce the same kind of exact duplication in the dataset,  because of syntax differences; 
There is a possibility of cross-language clones~\cite{perez2019cross}; the study of this
is left for future work. 


\takeaway{5}{Combining multilingual datasets is unlikely to cause exact duplication, because of syntax differences. More study
is needed to study the effect of cross-language clones.}

\subsection{Threats: Other metrics?}

Following \cxglue benchmark recommendation, we evaluate the code summarization task with smooth sentence BLEU-4~\cite{lin2004orange} throughout this paper. However, other recognized metrics are are available (\eg ROUGE-L~\cite{lin2004rouge}, METEOR~\cite{banerjee2005meteor}).
Prior works~\cite{shi2021neural,roy2021reassessing,gros2020code}  provide a careful analysis of the metrics, baselines, evaluations for code summarization task.
Table~\ref{met-rouge} shows ROUGE-L and METEOR data; we find that multilingual fine-tuning increases the overall performance by 4.89\% and 5.61\% in ROUGE-L and METEOR, respectively.    
As with BLEU-4, we find that multilingual fine-tuning shows similar performance gains with these metrics. We find 0.3\%-14.1\% improvement in ROUGE-L and 1.2\%-22.5\% gains in METEOR (except for PHP, were we see a 0.17\% \emph{decline}, not statistically significant).  We also see that Python shows the smallest improvement, not as strongly statistically significant. 
These metrics also indicate strong gains from multilingual training for low-resource and narrow-domain languages (\ie Ruby and JavaScript). 

\begin{table}[h]

\centering
\resizebox{\columnwidth}{!}{%
\renewcommand{\arraystretch}{1.2}

\begin{tabular}{lcccccccc}
\hline
\multicolumn{1}{c}{\multirow{2}{*}{Language}} & \multicolumn{4}{c}{ROUGE-L}                                                                                                                                                                                                                & \multicolumn{4}{c}{METEOR}                                                                                                                                                                                                                 \\ 
\multicolumn{1}{c}{}                          & \multicolumn{1}{c}{\mlcbert} & \multicolumn{1}{c}{Improve.*} & \multicolumn{1}{c}{\begin{tabular}[c]{@{}c@{}}Effect \\ Size\end{tabular}} & \multicolumn{1}{c}{\begin{tabular}[c]{@{}c@{}}p-value \\ (adjusted)\end{tabular}} & \multicolumn{1}{c}{\mlcbert} & \multicolumn{1}{c}{Improve.*} & \multicolumn{1}{c}{\begin{tabular}[c]{@{}c@{}}Effect \\ Size\end{tabular}} & \multicolumn{1}{c}{\begin{tabular}[c]{@{}c@{}}p-value \\ (adjusted)\end{tabular}} \\ \hline
Ruby                                            & 24.36                                    & +14.10\%                          & 0.087                                                                       & \textless{}0.001                                                                   & 21.96                                    & +22.54\%                         & 0.125                                                                       & \textless{}0.001                                                                   \\
JavaScript                                      & 24.30                                     & +7.05\%                          & 0.022                                                                       & \textless{}0.001                                                                   & 21.59                                    & +11.40\%                          & 0.030                                                                        & \textless{}0.001                                                                   \\
Java                                            & 34.89                                    & +3.32\%                          & 0.020                                                                        & \textless{}0.001                                                                   & 31.73                                    & +4.41\%                         & 0.020                                                                        & \textless{}0.001                                                                   \\
Go                                              & 37.36                                    & +2.69\%                          & 0.024                                                                       & \textless{}0.001                                                                   & 30.28                                    & +3.73\%                          & 0.023                                                                       & \textless{}0.001                                                                   \\
PHP                                             & 38.81                                    & +0.34\%                          & -8.65E-05                                                                   & 0.508                                                                              & 35.52                                    & -0.17\%                         & -0.003                                                                      & 0.779                                                                              \\
Python                                          & 32.86                                    & +1.86\% & 0.015                                                                       & \textless{}0.001                                                                   & 27.75                                    & +1.24\%                          & 0.004                                                                       & 0.033                                                                              \\ \hline
Overall                                                       & 32.10                                     & +4.89\%                          & \multirow{2}{*}{0.016}                                                      & \multirow{2}{*}{\textless{}0.001}                                                  & 28.14                                    & +5.61\%                          & \multirow{2}{*}{0.013}                                                      & \multirow{2}{*}{\textless{}0.001}                                                  \\
\begin{tabular}[c]{@{}l@{}}Overall \\ (weighted)\end{tabular} & 34.82                                    & +2.24\%                          &                                                                             &                                                                                    & 30.52                                    & +2.59\%                          &                                                                             &                                                                                      \\ \hline

\multicolumn{9}{l}{*Improvement reported over \cbert}

\end{tabular}
}

\vspace{0.05in}
\caption{{\em {\small Performance improvement in ROUGE-L and METEOR for code summarization task}}}
\label{met-rouge}

\vspace{-0.2in}

\end{table}

\takeaway{6}{We observe performance improvement in all code summarization metrics with multilingual fine-tuning.}

\subsection{Monolingual minibatches? or multilingual?}
While training deep neural networks with stochastic gradient descent,
gradients (multivariate derivatives of loss \emph{w.r.t} learnable parameters) are estimated
over \emph{mini-batches}, rather than calculating loss gradients over the entire training set; these estimates are used to adjust
the weights in the  network. Better choices of mini-batches could  improve convergence behavior. 
With multilingual training, a natural question arises: is it better to sequentially interperse \emph{monolingual} mini-batches
(\emph{e.g.,} first a Java minibatch, then Ruby minibatch and so on, before going back to Java?)
or should we make each minibatch \emph{per se} multilingual?

In the previous experiments, we had randomly sort the dataset; hence, our mini-batches are also multilingual. 
So we deliberately tried sequentially monolingual minibatching during multilingual fine-tuning. 
 We find that sequentially monolingual minibatch training appears to somewhat degrade performance: we observe the overall performance goes down by 1.05\%. However, the change is not statistically significant for any language. We omit the actual numerical details, for space reasons, since we didn't find any strong results in either direction. 

\takeaway{7}{We don't find any clear  difference between multilingual mini batches 
and (interspersed) monolingual mini batches.}

\subsection{Multilingual model as pre-trained model}

Our findings provide evidence supporting the claim that a multilingual fine-tuned model is effective for code summarization task, which outperforms all the models trained with monolingual datasets. Could this this improved  multilingual model further benefit from  a secondary,
\emph{monolingual} fine-tuning exercise, where it receives specialized fine-tuning for each language separately? 
To evaluate this intriguing and promising idea, we load the model with the weights from multilingual fine-tuning, and fine-tune it, again,  for each individual language. Table~\ref{pre-ablation} shows that
We found some minor performance improvement for JavaScript and Python. However, the performance goes down for other languages. 
We do not find evidence that a secondary, monolingual fine-tuning is helpful; further work is needed to understand
why, and perhaps develop other ways this idea might yield further improvement. 

\begin{table}[h]

\centering
\resizebox{\columnwidth}{!}{%
\renewcommand{\arraystretch}{1.2}

\begin{tabular}{llllll}
\hline
\multicolumn{1}{c}{Language}          & \multicolumn{1}{c}{\begin{tabular}[c]{@{}c@{}} \mlcbert\end{tabular}} & \multicolumn{1}{c}{\begin{tabular}[c]{@{}c@{}}\mlcbert \\as pre-training\end{tabular}} & \multicolumn{1}{c}{Improvement} & \multicolumn{1}{c}{\begin{tabular}[c]{@{}c@{}}Effect\\ Size\end{tabular}} & \multicolumn{1}{c}{\begin{tabular}[c]{@{}c@{}}p-value\\ (adjusted)\end{tabular}} \\ \hline
Ruby                                    & \textbf{14.75                                                                                             } & 14.58                                                                                                               & -1.15\%                          & -0.016  & 0.303                                                                             \\
JS                                      & 15.80                                                                                              & \textbf{16.47                                                                                                              } & +4.24\%                          & 0.024      & \textless{}0.001                                                                  \\
Java                                    & \textbf{20.11                                                                                             } & 19.81                                                                                                               & -1.49\%                          & -0.003                                                                     & 0.303                                                                             \\
Go                                      & \textbf{18.77                                                                                             } & 17.97                                                                                                               & -4.26\%                          &  -0.012 & \textless{}0.001                                                                  \\
Php                                     & \textbf{26.23}                                                                                              & 25.52                                                                                                               & -2.71\%                          & -0.017 & \textless{}0.001                                                                  \\
Python                                  & 18.71                                                                                              & \textbf{18.83                                                                                                              } & +0.64\%                          &  0.010 & \textless{}0.001                                                                  \\ \hline
\multicolumn{1}{l}{Overall}                                                      & \multicolumn{1}{l}{\textbf{19.06}}                                                                         & \multicolumn{1}{l}{18.86}                                                                                          & \multicolumn{1}{l}{-1.05\%}     & \multicolumn{1}{l}{\multirow{2}{*}{-0.003}}                               & \multicolumn{1}{l}{\multirow{2}{*}{0.005}}                                       \\ 
\multicolumn{1}{l}{\begin{tabular}[c]{@{}l@{}}Overall\\ (weighted)\end{tabular}} & \multicolumn{1}{l}{\textbf{20.74}}                                                                         & \multicolumn{1}{l}{20.43}                                                                                          & \multicolumn{1}{l}{-1.47\%}     & \multicolumn{1}{l}{}                                                      & \multicolumn{1}{l}{}                                                             \\ \hline
\end{tabular}
}

\vspace{0.05in}
\caption{{\em {\small Multilingual model as pre-trained model}}}
\label{pre-ablation}
\vspace{-0.3in}

\end{table}

\takeaway{8}{We don't find evidence that applying a secondary, mono-lingual fine-tuning provides benefits for all languages.}

%% file: related_work.tex
\section{Related work}

\noindent{\underline{\em Code summarization:}} 
Code summarization has recently been a hot topic. More than 30 papers have been published in the last five years that follow some form of encoder-decoder architecture~\cite{roy2021reassessing}. Several works~\cite{roy2021reassessing,shi2021neural,gros2020code} discuss the evaluations, metrics, and baselining. Roy \etal show that metric improvements of less than 2 points do not guarantee systematic improvements in summarization and are not reliable as proxies of human evaluation~\cite{roy2021reassessing}. We find more than 2 points of improvement for Ruby and almost 2 points of improvement for JavaScript. We observe less than 2 points in other languages. It should also be noted that we \emph{don't} use the corpus-level metrics which
Roy \etal show is problematic; we use pairwise comparisons on the test-sets. 
Finally, we note that progress in both code \& NLP 
occurs in small steps over decades, and innovations (especially ones that could cumulate with others) such as ours can be an important part of research community's long-term pursuit of practically relevant performance improvements. 

Pre-trained models~\cite{feng2020codebert,ahmad-etal-2021-unified,qi2021prophetnet,phan2021cotext,liu2019roberta} are proven to be more effective than prior models. Different pre-trained models are trained with the different pre-trained objectives even though fine-tuning steps are almost similar for all the models. As discussed earlier in Section~\ref{cgcmodels}, \cbert is an encoder model, pre-trained with MLM and Repace Token Detection objectives. Unlike \cbert, PLBART~\cite{ahmad-etal-2021-unified} is an encoder-decoder model which is trained as a denoising auto-encoder. Though all the models are pre-trained with different training objectives, there is one thing common among all the models: using Transformers as core architecture.  

Parvez \etal very recently present an approach that augments training data using relevant code or summaries retrieved from a database (\eg GitHub, Stack Overflow)~\cite{parvez2021retrieval}. They apply this approach  on 
monolingual Java and Python datasets from \cxglue and claim gains over \mlcbert \& \mlgcbert for code summarization. \emph{Prima  facie}, multilingual fine-tuning is 
complementary to their approach; this needs to be studied. 

\noindent{\underline{\em Code retrieval and method name prediction:}} 
Code retrieval is also getting attention recently. There are multiple datasets for this task. \cxglue introduces a monolingual python dataset (taken initially from CodeSearchNet) abstracting the function names and variables. Guo \etal modify the multilingual CodeSearchNet dataset and achieve state-of-the-art performance on this task. However, using multilingual training, we show that both \cbert and \gcbert can be improved. There is one other very recent paper, CLSEBERT~\cite{wang2021clsebert} which reports (in an unpublished, non-peer-reviewed report) better performance than us in all languages except Ruby. We could not show the effectiveness of multilingual training on CLSEBERT because the authors have not released the code implementation yet. Note that like code summarization, we focus only on the work using CodeSearchNet multilingual dataset. 

CodeSearchNet dataset can be easily adapted to method name prediction task. 
Several earlier works address method name prediction, in a Java-only such as Code2Seq~\cite{alon2018codeseq}, Allamanis~\cite{allamanis2016convolutional}. They all use a conventional
single-stage machine-learning approach (no pre-training + fine-tuning). Our goal here is to simply demonstrate that multilingual fine-tuning improves upon  monolingual fine-tuning for the method-naming task, so we demonstrate using \cbert. Our numbers are  roughly comparable with previously reported results, but we cannot make a precise comparison because of differences in subtokenization, and because our datasets are multilingual whereas previous work has largely been monolingual. We are simply arguing here our data suggests that multilingual fine-tuning is broadly beneficial in different tasks. 

It would certainly be interesting to use same-domain data for fine-tuning. For example, summarizing methods in Android apps might work better if trained on Android app corpora; however curated, domain-specific datasets for each domain are needed, and may not always be possible, depending on the domain. However, cross-language data is already available, and we show that it does indeed help improve performance! The effect of domain-specific corpora is left for future work.

%
%
  

%% file: conclusion.tex
\section{Conclusion}
We began this paper with three synergistic observations: \emph{First,} when solving the
\emph{same problem}, even in different programming languages, programmers are more likely
to use similar identifiers (than when solving \emph{different} problems). 
\emph{Second,} identifiers appear to be relatively much more important than syntax markers 
when training machine-learning models to 
perform code summarization. \emph{Third,} we find that quite often a model trained in one programming language
achieves surprisingly good performance on a test set in a different language, sometimes even surpassing a model trained on the same
language as the test set! 
Taken together, these findings suggest that pooling data across languages, thus creating
\emph{multilingual} training sets, could improve performance for any language, particularly perhaps
languages with limited resources, as has been found in Natural-language processing~\cite{dabre2020survey,ha2016toward,ranathunga2021neural,tang2020multilingual}.  We test this
theory, using two BERT-style models, \cbert, and \gcbert, with encouraging results.

Foundation models~\cite{bommasani2021opportunities}  are currently achieving best-in-class performance for a wide range of tasks
in both natural language and code. The models work in 2 stages, 
first ``pre-training'' to learn statistics of language (or code) construction from very large-scale corpora in a self-supervised
fashion, and then using smaller labeled datasets to ``fine-tune'' for specific tasks. 
We adopt the multilingual \cxglue dataset, and the pre-trained \cbert and \gcbert models, 
and study the value of multilingual fine-tuning for a variety of tasks. We find evidence suggesting that multilingual
fine-tuning is broadly beneficial in many settings. Our findings suggest that multilingual training
could provide added value in broad set of settings, and merits further study. 

\noindent{\em Acknowledgements:} This material is based upon work supported by the  U.S. National Science Foundation under Grant Nos. 1414172, and 2107592. Any opinions, findings, and conclusions or recommendations expressed in this material are those of the author(s) and do not necessarily reflect the views of the National Science Foundation. Ahmed was also supported by UC Davis College of Engineering Dean's Distinguished Fellowship.